\documentclass[11pt]{article}
\usepackage{thumbpdf,lmodern}
\usepackage{framed}
\usepackage{amsbsy} 
\usepackage[parfill]{parskip} 
\usepackage{graphicx} 
\graphicspath{ {.} }
\usepackage{fancyvrb}
\usepackage{float}
\usepackage[authoryear,round]{natbib}
\usepackage{authblk}  
\usepackage{listings} 
\title{Parsimoniously Fitting Large Multivariate Random Effects in \textbf{glmmTMB}}

\author[1]{Maeve McGillycuddy }
\author[2]{Gordana Popovic}
\author[3]{Benjamin M.  Bolker}
\author[1]{David I. Warton}
\affil[1]{School of Mathematics and Statistics,
  UNSW Sydney,  Australia}
\affil[2]{Stats Central, Mark Wainwright Analytical Centre,
  UNSW Sydney,  Australia}
  \affil[3]{Department of Mathematics \& Statistics, McMaster University, Canada}

\begin{document}

\maketitle

\begin{abstract}
Multivariate random effects with unstructured variance-covariance matrices of large dimensions, $q$, can be a major challenge to estimate. In this paper, we introduce a new implementation of a reduced-rank approach to fit large dimensional multivariate random effects by writing them as a linear combination of $d < q$ latent variables. By adding reduced-rank functionality to the package \texttt{glmmTMB}, we enhance the mixed models available to include random effects of dimensions that were previously not possible. We apply the reduced-rank random effect to two examples, estimating a generalized latent variable model for multivariate abundance data and a random-slopes model. 
\end{abstract}

\section{Introduction} \label{sec:intro}

When fitting a mixed effects model, it is often necessary to use a multivariate random effect with a non-diagonal covariance matrix in order to introduce sets of correlated parameters to a model. This approach is needed, for example, when fitting random-slopes models \citep{bolker2009generalized, asar2020lmm}, to account for correlation between the slope coefficient(s) and intercept terms, and when using random effects to induce correlation in multivariate data \cite[for example]{coull2000random,pollock2014jsdm}. Without imposing structure on the variance-covariance matrix, the number of parameters that need to be estimated increases quadratically with the dimension of the random effect (specifically, $q(q+1)/2$ parameters need to be estimated for an unstructured $q \times q$ covariance matrix), and estimation quickly becomes challenging as $q$ gets larger.

For example, in Section \ref{sec:wf_data} we describe a study of the effect of wind farms on fish assemblages, where we count individuals of multiple species at several sites. We wish to use a multivariate random effect to estimate correlation across species. We can do this using a mixed model fitted using the \texttt{lme4} package \citep{bates2014fitting} in \texttt{R} \citep{R} as follows:
\begin{verbatim}
R> glmer(abundance ~  Zone + Year + (Species + 0 | ID), 
+    family = "poisson", data = windfarm)
\end{verbatim}
or equivalently, using the \texttt{glmmTMB} package \citep{brooks2017glmmtmb}:
\begin{verbatim}
R> glmmTMB(abundance ~  Zone + Year + (Species + 0 | ID), 
+    family = "poisson", data = windfarm)
\end{verbatim}
We show in Appendix \ref{app:summary} that if there are only two species in the data set then this approach is reasonable (and these two lines of code produce identical answers, up to machine error), but convergence issues start to be seen when there are three or more species. The complete data set that we wish to analyse has nine species, and similar types of data frequently involve many more species, sometimes thousands \citep{niku2019gllvm}.

One common way to deal with high dimensionality is to use a reduced-rank approach, making simplifying assumptions that reduce the dimension of the problem to $d < q$. Reduced-rank approaches have  seen considerable use in bioinformatics (e.g., \citealp{smith2001analyzing, buettner2015computational}) and spatial statistics \citep{cressie2008fixed, banerjee2008GPP}. A reduced-rank approach to fitting a multivariate random effect involves writing it as a linear combination of $d$ latent variables, often referred to as a factor analytical model \citep{bartholomew2011latent}; in the case of exponential family responses, it is sometimes called a generalized latent variable model  \cite[GLVM: ][]{skrondal2004generalized}. 
 
GLVMs have been used frequently in ecology and the social sciences, early examples being models of the presence-absence of fish species \citep{walker2011random} and of polytomous party choice and rankings data \citep{skrondal2003poly}. GLVMs can be technically challenging to fit, but there are a number of dedicated software solutions. The \texttt{gllamm} package \citep{rabe2004generalized} in \texttt{Stata} \citep{Stata} uses adaptive Gaussian quadrature to integrate out the latent variables. Early software written in \texttt{R}, with ecologists in mind, used Bayesian Markov Chain Monte Carlo \citep{HuiBoralPackage,OvaskainenHMSCpackage}. Substantially faster fits can be obtained using a Laplace \citep{huber2004estimation, niku2017generalized} or variational approximation \citep{HuietalVariationalLVM} to the marginal likelihood, as implemented in the \texttt{R} package \texttt{gllvm} \citep{niku2019gllvm}. These tools were written with a particular model in mind, where a multivariate random intercept is used to induce correlation across many responses, and fixed effects are used to relate the linear predictor to measured variables, although there have been some extensions to relax this constraint in different ways \citep{van2021model, niku2021analyzing}. However, this software is unable to handle a range of common sampling designs. An example considered in Section~\ref{sec:Implementation} is when the multivariate random effect is not an intercept term.

In this paper we add reduced-rank functionality to the \texttt{glmmTMB} package \citep{brooks2017glmmtmb} for flexible mixed effects models that can include factor analytic terms, for multivariate random effects of large dimension. The \texttt{glmmTMB} package is built as an extension to \texttt{lme4} \citep{bates2014fitting}, widely used in the applied sciences for problems involving a range of study designs, including multi-level and repeated measures designs.  The \texttt{glmmTMB} package uses a similar interface to \texttt{lme4} but exploits automatic differentiation for faster estimation of mixed effects models \citep{brooks2017glmmtmb}. By adding reduced-rank functionality to \texttt{glmmTMB}, we enrich the class of mixed models that can be fitted to include multivariate random effects with much larger dimension than was previously possible, such that we can now routinely fit random effects of dimension in the hundreds, or perhaps in the thousands. This new class of models includes, for example, GLVMs with functionality for any of the study designs that can be analysed using \texttt{lme4}, including multi-level or repeated measures designs.

Section \ref{sec:models} provides an overview of the generalized linear mixed model, the factor analytic extensions to handle multivariate random effects of high dimension, and the estimation approach used in \texttt{glmmTMB} to fit such models. We describe the usage of  \texttt{glmmTMB} to fit reduced-rank multivariate random effects. We analyse two different data sets, from ecology and the social sciences, in Section \ref{sec:Application}. Section \ref{sec:Conclusion} concludes the paper.

\section{Methods} \label{sec:models}

We start by introducing a generalized linear mixed model and the factor analytic variant we use to handle multivariate random effects with large dimension. Then we discuss the estimation process for these models and the interface to fit reduced-rank multivariate random effects as implemented in \texttt{glmmTMB}.

\subsection{Models}
Let $y_{ij}$ be the response for $i = 1, \ldots, n_{j}$ observational units in cluster  $j = 1, \ldots, m$. A vector of $p$ fixed effect covariates, $\boldsymbol{x}_{ij}$, and $q$ random effect covariates, $\boldsymbol{z}_{ij}$, may also be recorded for each unit. For a generalized linear mixed model (GLMM), conditional on the vector of random effects, $\boldsymbol{b}_j$, and the vector of parameters, $\boldsymbol{\Psi}$ (defined below), the responses are assumed to come from the exponential family of distributions, $f( y_{ij} | \boldsymbol{b}_{j}, \boldsymbol{\Psi} ) = \exp[ (y_{ij} a( \eta_{ij} ) - c(\eta_{ij}) )/ \phi + d(y_{ij}; \phi)]$ where $a(\cdot)$, $c(\cdot)$ and $d(\cdot)$ are known functions that depend on the chosen distribution $f$, $\eta_{ij}$ are canonical parameters, and $\phi$ is a dispersion parameter.  Then the mean response, denoted as $\mu_{ij}$, regressed against the fixed and random covariates can be specified as
\begin{equation}\label{model_eqn}
\eta_{ij} = g(\mu_{ij}) = \boldsymbol{x}_{ij}^\top\boldsymbol{\beta}+ \boldsymbol{z}_{ij}^\top\boldsymbol{b}_j
\end{equation}
where $g(.)$ is the link function, $\boldsymbol{\beta}$ is a $p$-dimensional vector of regression coefficients related to the covariates, $\boldsymbol{x}_{ij}^\top = (1, x_{1ij}, \ldots, x_{pij})$, and $\boldsymbol{z}_{ij}^\top = (1, z_{1ij}, \ldots, z_{qij})$ is the vector of random effect covariates. The unconditional distribution of the random effects, or cluster level errors, $\boldsymbol{b}_{j}$,  is assumed to follow a multivariate normal distribution with mean zero and a parameterized $q \times q$ variance-covariance matrix, $\boldsymbol{\Sigma}$, i.e., $ \boldsymbol{b}_{j} \sim \mathcal{N}(\boldsymbol{0}, \boldsymbol{\Sigma})$. The variance-covariance matrix, $\boldsymbol{\Sigma}$, controls the variances of  and correlations between units in clusters. The most flexible option for $\boldsymbol{\Sigma}$ is an unstructured variance-covariance matrix, which requires $q(q+1)/2$ parameters. For models with large multivariate random effects, this flexibility becomes a problem, with the number of parameters in $\boldsymbol{\Sigma}$ increasing quadratically with the size of the random effect, $q$.

A reduced-rank approach to fit a multivariate random effect involves expressing it as a linear function of $d$ latent variables: 
\begin{equation}
\boldsymbol{b}_{j} = \boldsymbol{\Lambda} \boldsymbol{u}_{j}
\end{equation}
where $\boldsymbol{u}_{j}$ is a vector of $d$ latent variables, each of which is independent and standard normal, and $\boldsymbol{\Lambda}$ is a $q \times d$ matrix of factor loadings. The latent variables have a zero mean and unit variance, without loss of generality. Upper triangular elements of $\boldsymbol{\Lambda}$ are set to zero to assist with parameter identifiability, without loss of generality. Finally, we let $\boldsymbol{\Psi} = \left\{ \boldsymbol{\beta}, \boldsymbol{\Lambda}, \phi  \right\}$ denote the complete set of model parameters.

Given these definitions, the multivariate random effect is distributed as 
\begin{equation}\label{bj_dist}
\boldsymbol{b}_j \sim \mathcal{N}\left( \boldsymbol{0}, \boldsymbol{\Lambda \Lambda^\top} \right)
\end{equation}
which is a reduced-rank approximation of $\boldsymbol{b}_j \sim \mathcal{N}(\boldsymbol{0}, \boldsymbol{\Sigma})$, as often seen in factor analytic models \citep{bartholomew2011latent, niku2017generalized}. This approach makes it possible to fit large multivariate random effects, because the number of parameters required in the variance-covariance matrix of $\boldsymbol{b}_j$ is now $dq-{d\choose 2}$, which (for fixed $d$) increases only linearly with $q$.
\subsection{Estimation} 

Conditional on the latent variables, $\boldsymbol{u}_{j}$, responses are assumed to be independent, hence $f(\boldsymbol{ y_{j}} | \boldsymbol{u}_{j}, \boldsymbol{\Psi}) = \prod_{i=1}^{n} f(y_{ij} | \boldsymbol{u}_{j}, \boldsymbol{\Psi}) $. As the latent variables are not observed, they are integrated out, leading to the marginal log-likelihood:
\begin{equation} \label{marg_ll}
l(\boldsymbol{\Psi}) = \sum_{j=1}^{m}  \log( f(\boldsymbol{ y}_{j}, \boldsymbol{\Psi})) = \sum_{j=1}^{m}  \log \left( \int \prod_{i=1}^{n}  f(y_{ij} | \boldsymbol{\Psi}, \boldsymbol{u}_{j}) f(\boldsymbol{u}_{j}) d\boldsymbol{u}_{j} \right).
\end{equation}

In some cases this expression can be explicitly solved and expressed in closed form, but for non-normal distributions it does not generally have a closed form.  A number of estimation methods have been proposed to approximate the marginal likelihood including Laplace approximation, numerical integration using adaptive quadrature \citep{skrondal2004generalized}, Monte Carlo integration \citep{hui2015} and more recently variational approximation \citep{HuietalVariationalLVM}.

We focus on the Laplace approximation of the marginal likelihood, which is widely used for GLMMs \citep{raudenbush2000maximum} as well as GLVMs \citep{huber2004estimation,niku2017generalized}. By writing Equation \ref{marg_ll} in the form $l(\boldsymbol{\Psi}) = \sum_{j=1}^{m}  \log \int \exp(m Q(\boldsymbol{y_{j}}, \boldsymbol{\Psi}, \boldsymbol{u}_{j})) d\boldsymbol{u}_{j}  $, where
\begin{equation}
 Q(\boldsymbol{y}_{j}, \boldsymbol{\Psi}, \boldsymbol{u}_{j}) = \frac{1}{n} \left[ \sum_{i=1}^{n} \left\{ \frac{y_{ij} a( \eta_{ij} ) - c(\eta_{ij})}{\phi} + d(y_{ij}; \phi) \right\} - \frac{\boldsymbol{u}_{j}^\top\boldsymbol{u}_{j}}{2} - \frac{q}{2} \log (2\pi) \right]
\end{equation}
we can apply Laplace's method for integral approximation around its mode, $\boldsymbol{u}_{j}$. Assuming $\boldsymbol{\hat{u}}_{j}$ maximises $Q(\boldsymbol{y}_{j}, \boldsymbol{\Psi}, \boldsymbol{u}_{j})$, the approximation is derived by expanding $Q(\boldsymbol{y}_{j}, \boldsymbol{\Psi}, \boldsymbol{{u}}_{j})$ as a second order Taylor series around the mode, $\boldsymbol{\hat{u}}_{j}$. Following  \cite{huber2004estimation}, a Laplace approximation of the marginal log-likelihood function of the GLVM defined in Equation \ref{model_eqn} can be written as
\begin{equation}
    l(\boldsymbol{\Psi}) = \sum_{j=1}^{m} \left( -\frac{1}{2} \log \det \left\{ G(\boldsymbol{\Psi},\boldsymbol{\hat{u}}_j) \right\} + \sum_{i=1}^{n}\left\{ \frac{y_{ij} a( \eta_{ij} ) - c(\eta_{ij})}{\phi}   + d(y_{ij}; \phi) \right\} - \frac{1}{2}\boldsymbol{\hat{u}}_j^\top\boldsymbol{\hat{u}}_j \right)
\end{equation}
where 
\begin{equation}
     G(\boldsymbol{\Psi},\boldsymbol{\hat{u}}_j) = 
     \frac{\partial^2  -Q(\boldsymbol{y}_{j}, \boldsymbol{\Psi}, \boldsymbol{u}_{j})}{\partial \boldsymbol{u}_{j}^\top \partial \boldsymbol{u}_{j}} = \sum_{i=1}^{n} \frac{1}{\phi} \frac{\partial^2 \{ -y_{ij} a( \eta_{ij} ) + c(\eta_{ij})\} }{{\partial \boldsymbol{u}_j^\top \partial \boldsymbol{u}_j }} \bigg|_{\boldsymbol{u}_j = \boldsymbol{\hat{u}}_j}   + I_q
\end{equation}
and $\boldsymbol{\hat{u}}_j$ is the maximum of $Q(\boldsymbol{y}_{j}, \boldsymbol{\Psi}, \boldsymbol{u}_{j})$.

In \texttt{glmmTMB} we maximise a Laplace approximation of the marginal log-likelihood obtained from the package Template Model Builder (\texttt{TMB}) \citep{kristensen2015tmb} in \texttt{R} (or more precisely, minimise the negative log-likelihood). \texttt{TMB} evaluates the Laplace approximation and its derivatives using automatic differentiation. Any gradient-based optimization method available in \texttt{R} can be used to do the maximization, by default \texttt{glmmTMB} uses \texttt{nlminb()}.  \texttt{TMB} then uses the generalised delta method to calculate marginal standard deviations of fixed and random effects \citep{kass1989approximate}.

\subsection{Software interface} \label{sec:Implementation}

A reduced rank covariance structure is specified in \texttt{glmmTMB} using \texttt{rr}. For example, suppose \texttt{x} is a matrix of predictors with $p$ columns, and that we want to apply $p$-dimensional random coefficients that take different values for different levels of a grouping variable \texttt{group} to predictors \texttt{x}. To specify that these random coefficients are drawn from a multivariate normal distribution whose variance-covariance matrix has rank two (that is, they can be written as a linear combination of two independent latent variables) we use the \texttt{rr} random effect structure in the formula as follows
\begin{verbatim}
rr(x | group, 2)
\end{verbatim}
So for example to model the mean of a response (\texttt{y}) as a function of \texttt{x} and let the coefficients of \texttt{x} vary randomly among groups as a function of two latent variables, the formula is
\begin{verbatim}
y ~ x + rr(x | group, 2)
\end{verbatim}

The non-negative integer after the comma in the formula specifies the number of latent variables, or rank, of the variance-covariance matrix of the  multivariate random effects, which defaults to $d=2$.The choice of the number of latent variables can be seen as a model selection problem \citep{hui2015}. Model selection tools including cross-validation, or information criteria can be used to select the rank \citep{bartholomew2011latent}.

One issue with fitting a factor-analytic model is that the likelihood function is often multimodal, which may lead to convergence to a local maximum. To overcome this we include a data-driven method based on the work of \cite{niku2019gllvm} to initialise values for parameters so that the maximising algorithm starts closer to the global maximum. The default in \texttt{glmmTMB} sets parameter values to zero, or one for fixed-effect parameters for some link functions. To control the algorithm used for initialising parameters a \texttt{start\_method} argument, specified as a list with components \texttt{method} and \texttt{jitter.sd}, has been added to \texttt{glmmTMBControl()}. Setting \texttt{method = "res"} fits a generalized linear model (GLM) to the data to obtain estimates of the fixed parameters, which are then used as starting values of the fixed parameters. From the fitted model, residuals are calculated for models using the Poisson, negative binomial, and binomial families using the method due to \cite{dunn1996randomized}, while the internal function \texttt{dev.residuals} is used for other families. Starting values for latent variables, $\boldsymbol{u}_{j}$, and factor loadings, $\boldsymbol{\Lambda}$, are obtained by applying a reduced-rank model to the residuals from the fitted GLM. To check stability of a solution, we suggest repeating a fit multiple times and adding variation to the starting values of latent variables when \texttt{method = "res"} by setting \texttt{jitter.sd = 0.2}, or similar, to jitter starting values for $\boldsymbol{u}_j$ by a normal variate with mean zero and standard deviation \texttt{jitter.sd}.

More examples of implementing the \texttt{rr} structure are shown in Section \ref{sec:Application}. Information on the reduced-rank and other available variance-covariance structures in \texttt{glmmTMB} can be found in \texttt{vignette("covstruct", package = "glmmTMB")}.

\section{Application} \label{sec:Application}

We present two applications to illustrate the use of a reduced-rank covariance structure in \texttt{glmmTMB}. The first example, using wind farm data, fits a generalized latent variable model to multivariate abundance data, a form of data often gathered for ecological studies. The second example presents a random-slopes model, with many random slopes, applied to data from the Progress in International Reading Literacy Study (PIRLS). 

\subsection{Wind farm data}\label{sec:wf_data}

\begin{figure}[tbp]
\includegraphics[width=\textwidth]{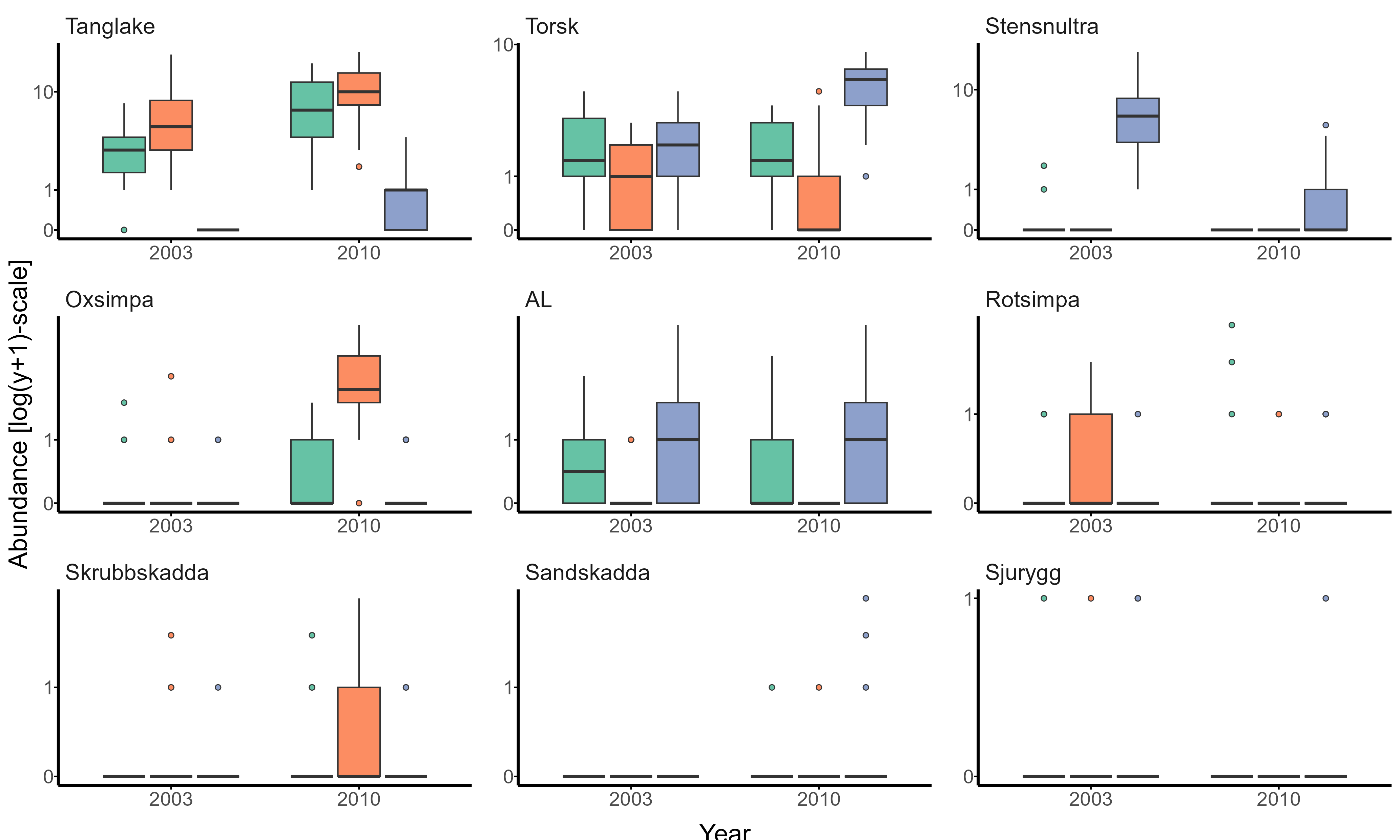}
\caption{Boxplot of fish abundance ($\log(y + 1)$ scale) for each species at three zones, windfarm (WF, green), north (N, orange), and south (S, lilac), before (2003) and after (2010) construction of the offshore wind farm.}
\label{fig:wfboxplot}
\end{figure}

The wind farm data were gathered for the Lillgrund offshore wind farm off the southern coast of Sweden, to study the effects of the wind farm on the demersal fish community \citep{bergstrom2013effects}. This is one of the few large-scale studies assessing the effects of offshore wind farms over a long period of time, and on more than one individual fish species. This study exemplifies a BACI (before-after-control-impact) design; abundance of fish was measured before (2003) and after construction (2010), in the wind farm zone and in two reference zones (southern and northern reference zones). Sampling occurred at fishing stations within each zone; stations remained the same throughout the study. Stations which were not sampled in both time periods were omitted. The raw abundance data (Figure \ref{fig:wfboxplot}) does not show an obvious windfarm effect, although species \texttt{Strensnultra} and \texttt{Oxsimpa} may show before-after differences for the south and north zones respectively. Because we are interested in the effect of wind farms, which were established between the sampling times, our interest is in the interaction between zone and year.

A multivariate random effect is required to account for correlation in species responses within a sample -- we expect correlation across species due to inter-specific interactions, or response to unobserved environmental conditions \citep{warton2015} and we wish to be able to estimate positive or negative correlations. We do not have any \emph{a priori} structure for this correlation, and for nine species we would require 45 parameters to estimate the variance-covariance matrix among species. Thus we would expect considerable instability in the fit, especially for correlation terms involving rarer species. In comparison, a reduced-rank covariance structure of rank two would require only 17 parameters.

We model abundances, $y_{ijk}$, as conditionally Poisson with mean $\mu_{ijk}$,  observed at $i = 1, \ldots, n$ samples, $k = 1,\ldots, l$ stations, for $j = 1, \ldots, p$ species such that:
\begin{equation}\label{eq:wfrr}
g(\mu_{ijk}) = \boldsymbol{x}_i^\top\boldsymbol{\beta} +   \boldsymbol{x}_i^\top\boldsymbol{b}_{j}^{[x]} + {b_{k}^{[s]}} + b_{ijk}^{[rr]}
\end{equation}

where $g(.)$ is the link function, $\boldsymbol{x}_{i}$ are vectors of environmental covariates specifying the intercept, zone, year and the interaction of zone and year, and $\boldsymbol{\beta}$ is a vector of fixed coefficients. We include a multivariate random effect (with $q=4$) on the environmental variables, $\boldsymbol{b}_{j}^{[x]}  \sim \mathcal{N}(\boldsymbol{0}, \boldsymbol{\sigma}_x^2\boldsymbol{I})$ to allow the effects of each covariate to vary across species. The random intercept for station, ${b_k^{[s]} \sim \mathcal{N}(0, \sigma^2_s)}$ is intended to account for paired sampling at stations, with data collected at two time points for each station. We assume each of these random effects is independent of all others and of the response (conditional on $\mu_{ijk}$). The correlation 
between species is induced by the reduced-rank random effect, $b_{ijk}^{[rr]}$, assumed to satisfy

\begin{equation}\label{eq:fa}
b_{ijk}^{[rr]} = \boldsymbol{\lambda}_j \boldsymbol{u}_{ik}
\end{equation}

where $\boldsymbol{u}_{ik}$ is a pair (dimension $d=2$) of latent variables, and the vector $\boldsymbol{\lambda}_j$ contains the corresponding factor loadings.

This model can then be fitted using the following command:

\begin{verbatim} 
R> glmmTMB(abundance ~ Zone * Year + diag(Zone * Year | Species) +
+    (1 | Station) + rr(Species + 0 | ID, 2), family = "poisson",
+    control = glmmTMBControl(start_method = list(method = "res")),
+    data = windfarm)
\end{verbatim}

The intercept was excluded from the reduced-rank random effect term (using \texttt{Species + 0} as the varying term) in order to aid interpretability of the correlation matrix discussed below. The estimated correlation matrix of the reduced-rank multivariate random effect can be obtained from the output of the \texttt{summary} method, or using the \texttt{VarCorr} method, in the same way that the estimated values for any variance-covariance matrix are returned by \texttt{glmmTMB}. 

The estimated correlation matrix of the random intercept for the wind farm data, using the rank-two model specified above, is as follows:

\begin{Verbatim}[fontsize=\tiny]
Conditional model:
 Groups  Name                Std.Dev. Corr                                       
 ID      SpeciesTanglake     0.4741                                                   
         SpeciesTorsk        0.1618    0.40                                           
         SpeciesStensnultra  0.6963    0.36  1.00                                     
         SpeciesOxsimpa      0.2092   -1.00 -0.34 -0.29                               
         SpeciesAL           0.5502   -0.85  0.13  0.18  0.89                         
         SpeciesRotsimpa     0.8684    0.24  0.99  0.99 -0.17  0.30                   
         SpeciesSkrubbskadda 0.6541    0.43  1.00  1.00 -0.36  0.10  0.98             
         SpeciesSandskadda   1.6886    0.39  1.00  1.00 -0.32  0.14  0.99  1.00       
         SpeciesSjurygg      0.6395    0.52  0.99  0.98 -0.46  0.00  0.95  0.99  0.99
\end{Verbatim}

\normalsize
These correlations are residual correlations between species not accounted for by the covariates and other random effects in the model. For example, the correlation between species \texttt{Oxsimpa} and \texttt{AL} is $0.89$ after controlling for the fixed and random covariates in the model. Note that these correlations are on the linear predictor scale (in this case the log scale), and the actual correlation observed in data is much weaker than these values, because of Poisson noise introduced by Equation \ref{eq:wfrr}. The marginal correlation structure is singular  (for $d < p$) as it is approximated from Equation \ref{bj_dist}, where $\boldsymbol{\Lambda}$ is reduced rank; this is not problematic as the estimates of the factor loadings, $\boldsymbol{\lambda}_j$, and latent variables, $\boldsymbol{u}_i$, are used in further analysis. Attempting to instead fit this model using an unstructured covariance structure runs into convergence problems; the data are of insufficient quality to support estimation of 45 separate variance parameters. 

\begin{figure}[tbp]
 \includegraphics[width=\textwidth]{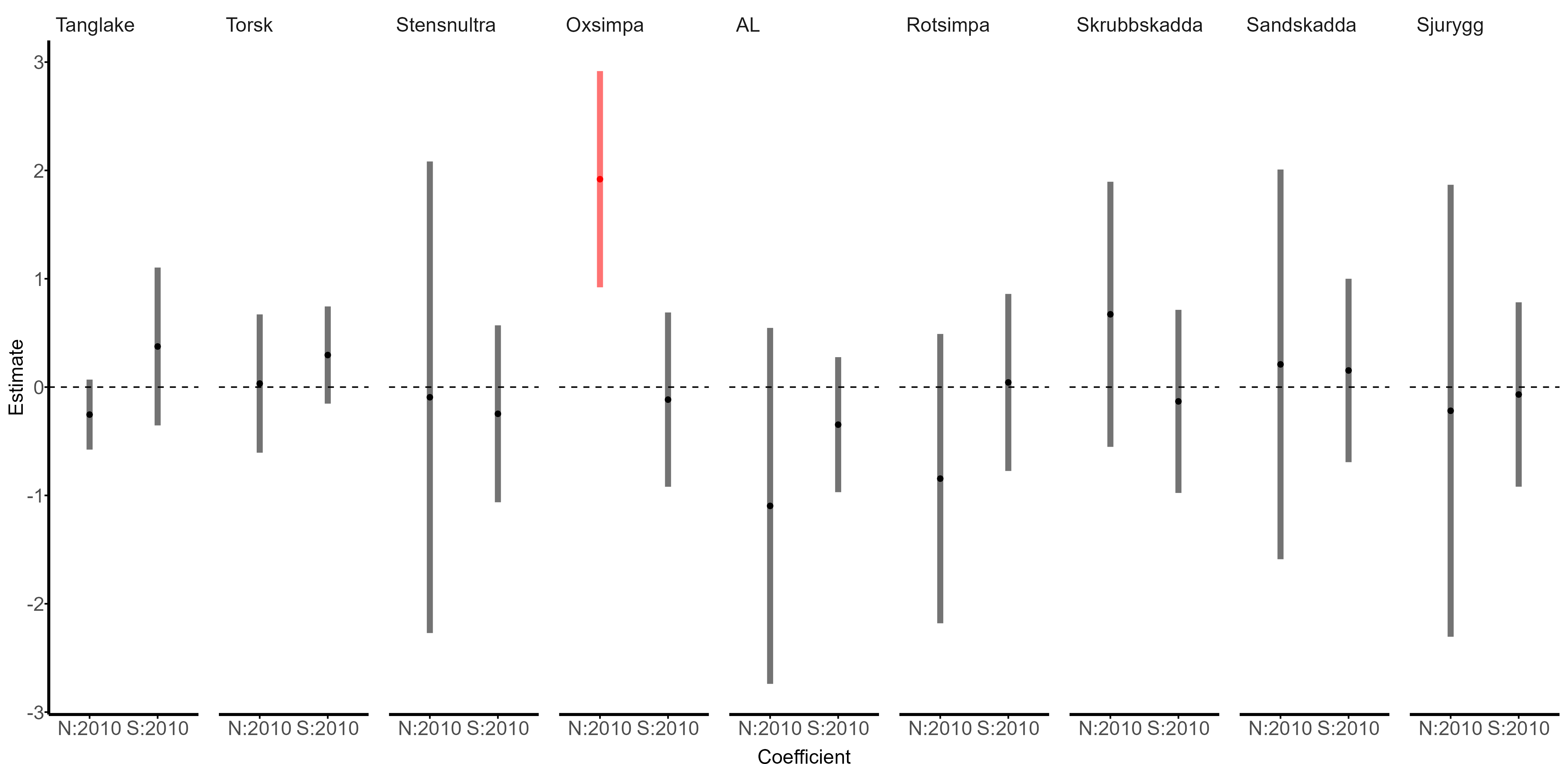}
 \caption{Conditional estimates (95\% confidence interval) of the \texttt{Zone} by \texttt{Year} interaction terms for species from the diagonal random effect. The contrast  between a zone (North, or South), and the Wind Farm zone in 2010 is shown.}
\label{fig:wfre}
\end{figure}

To test if the construction of an offshore wind farm had a significant effect on fish abundance, a parametric bootstrap analysis was conducted to test the \texttt{Zone} by \texttt{Year} interaction in both the fixed and random effect terms (code provided in Appendix \ref{sec:par}). A parametric bootstrap analysis was chosen over asymptotic  tests such as Wald and likelihood ratio tests, because the asymptotic distributions of these tests are usually unknown for mixed models \citep{bolker2009generalized}. The interactions were significant (LR = 27.35, $p = 0.001$), indicating that for at least one of the species, mean abundance across zones has changed (in a relative sense) since construction of the wind farm (Figure \ref{fig:wfre}). The estimate for Oxsimpa is clearly positive for the North-Wind Farm contrast in 2010 \citep{dushoff2019can}, while controlling for other covariates, although no such effect was seen for the South-Wind Farm contrast in 2010. Judging from Figure~\ref{fig:wfboxplot}, this effect probably has more to do with an increase in Oxsimpa in the North Zone, rather than an effect of wind farms.
\begin{figure}[tbp]
\includegraphics[width=\textwidth]{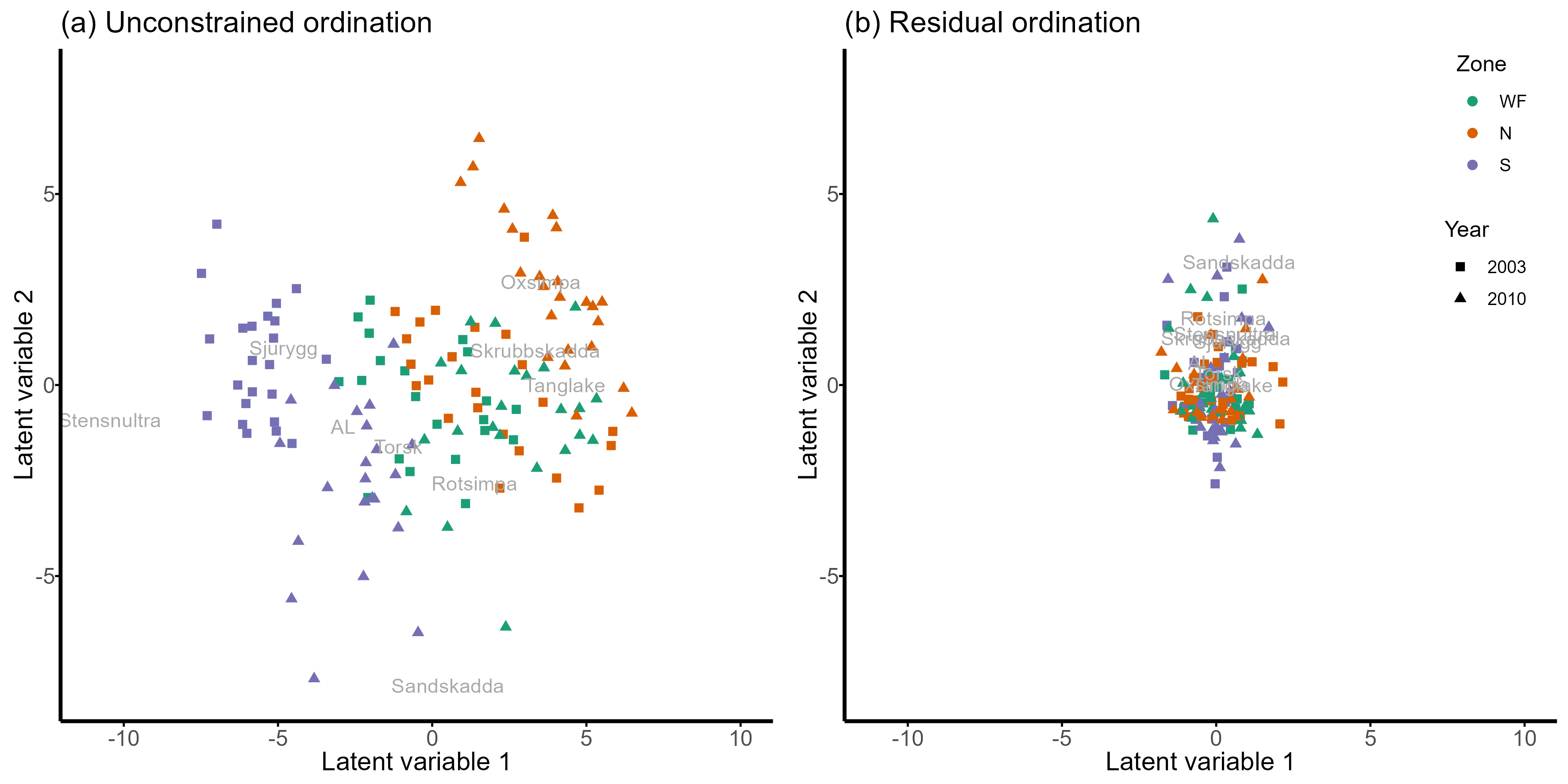}
\caption{Ordination biplot of the wind farm data (a) for the unconstrained model, (b) after including zone, year and the interaction in the model. Zones are shown in colours, year in symbols and species factor loadings are labelled accordingly.}
\label{fig:wford}
\end{figure}

To visualise the correlations between species, an ordination biplot can be produced from the estimated latent variables and factor loadings (Figure \ref{fig:wford}). An unconstrained ordination biplot (Figure \ref{fig:wford}a), plotting the latent variables from a model without and fixed effects predictors, shows a separation in the latent variables by \texttt{Zone} and \texttt{Year}, emphasising the importance of these variables. Adding factor loadings to the plot shows us how species vary across sites, with higher abundance of a species tending to be found in sites in the same direction as the species, with respect to the origin. Oxsimpa for example could be expected to be found in high numbers in the North Zone in 2010, and Stensnultra could expect to be found in high numbers in the South Zone, especially in 2003. Figure~\ref{fig:wfboxplot} corroborates both of these results. The relative positions of the species also gives information about their correlation --- because Oxsimpa and Stensnultra are negatively correlated they appear far from the origin but at opposite sides of the ordination, whereas the positively correlated Oxsimpa and Skrubbskadda are neighbours in the ordination plot.

After fitting the model in Equation~\ref{eq:wfrr}, which controls for the effects of \texttt{Zone}, \texttt{Year}, their interaction, and \texttt{Station}, the clustering patterns  by zone and sampling time disappear (Figure \ref{fig:wford}b). The points lie much closer together, with smaller loadings, reflecting the fact that adding predictors to the model substantially reduces the magnitude of the variance-covariance terms. This verifies that the prevailing patterns seen in the unconstrained ordination, and hence in the fish communities being sampled, can be explained by where and when samples were taken.

\subsection{PIRLS data}

Progress in international reading literacy study (PIRLS) is a large-scale international research project measuring reading literacy in children aged nine to ten years \citep{pirlsmethods}. PIRLS has been conducted every five years since 2001, with 61 countries participating in PIRLS 2016.  Published studies have proposed that school variables are more important than family  background in determining academic achievement in developing countries \citep{heyneman1982influences}. However, more recent studies report conflicting results which show that these variables are similar across countries \citep{maroco2021makes}, with authors proposing this homogenization may be due to the increase of mass schooling. Therefore, we are interested in exploring how the effect of school variables on literacy scores of students vary by country.

We propose that students from economically disadvantaged schools (\texttt{Eco\_disad}) with a library containing more books (\texttt{Size\_lib}) have higher literacy scores than students from a school with no library, but the difference in literacy scores will be less when students are from schools of higher economic background i.e., there is an interaction between the two school variables. Further, we would like to know how the interactive effect of these school-level variables will vary by country, hence we want to fit a random slopes model, with different \texttt{Eco\_disad:Size\_lib} effects for each country. Both \texttt{Eco\_disad} and \texttt{Size\_lib} are categorical variables with four levels, hence we need 15 parameters to characterise their joint effect and a 15-dimensional random effect in the model. Estimating an unstructured variance-covariance matrix $\boldsymbol{\Sigma}$ would require 136 parameters. In contrast, for a reduced-rank covariance structure of rank three, only 42 parameters are required -- AIC was used to select the rank of three.

We consider the model for literacy score, $y_{ijk}$, for student $i = 1, \ldots, n$, in school $j = 1,\ldots, m$ and country $k = 1, \ldots, p$ as follows:
\begin{equation}
y_{ijk} = \boldsymbol{x}_j^\top\boldsymbol{\beta} +  {b_{j}^{[s]}} + \boldsymbol{x}_j^\top\boldsymbol{b}_{k}^{[rr]} + \epsilon_{ijk}
\end{equation}
where $\boldsymbol{x}_{j}$ are vectors of covariates relating to school factors and $\boldsymbol{\beta}$ is the vector of fixed coefficients.  The random intercept for school,  $b_{j}^{[s]} \sim \mathcal{N}(0, \sigma^2_s)$, accounts for heterogeneity of average literacy scores between schools. The random intercept and slopes of school variables $\boldsymbol{b}_{k}^{[rr]} \sim \mathcal{N}(\boldsymbol{0}, \boldsymbol{\Lambda \Lambda^\top})$ allow the school variables to vary by country where $\boldsymbol{\Lambda}$ is the full matrix of factor loadings, and $\epsilon_{ijk} \sim \mathcal{N}(0, \sigma^2)$ is the residual error. 

This model can be fitted in \texttt{glmmTMB} using the following command:
\begin{verbatim}
R> glmmTMB(Overall ~  Size_lib * Eco_disad + (1 | School) +
+    rr(Size_lib * Eco_disad | Country, d = 3),
+    control = glmmTMBControl(start_method = list(method = "res")),
+    family = gaussian(),
+    data = pirls)
\end{verbatim}
The starting algorithm for initialising parameters specified in the \texttt{control} argument is needed when fitting this model, otherwise there will be convergence issues. 

\begin{figure}[tbp]
\includegraphics[width=\textwidth]{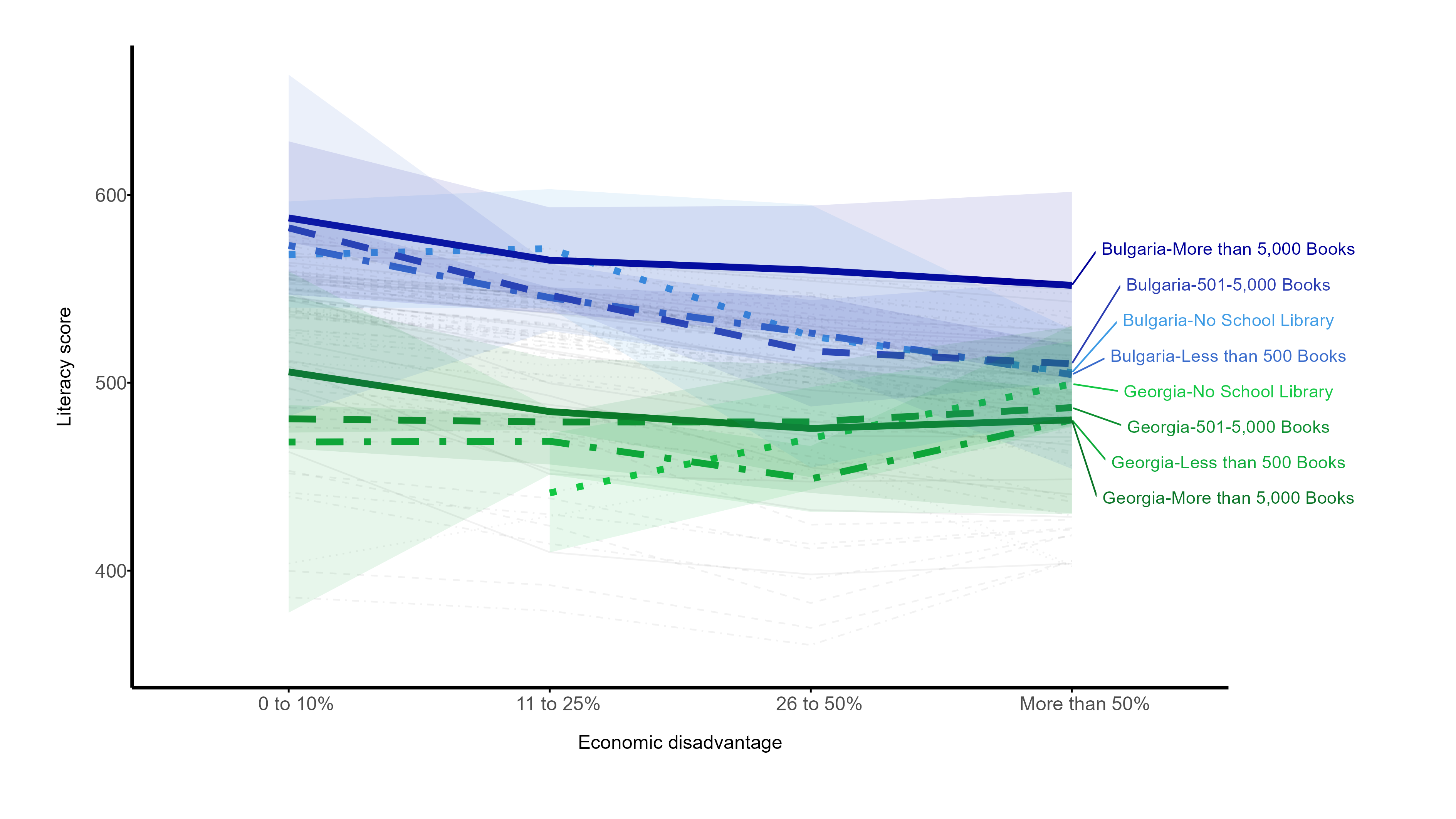}
\caption{Plot of average literacy score (95\% confidence interval shown by shaded area) of students in schools of different economic backgrounds and with varying library sizes in Georgia (green), Bulgaria (blue) and remaining countries (grey).}
\label{fig:pirls_int}
\end{figure}

The conditional estimates of the school random effects by country are  complex (Figure \ref{fig:pirls_re}): we will focus on a particular contrast between Bulgaria and Georgia (coloured results in Figure~\ref{fig:pirls_re}). This comparison is of interest because these two countries appear to have different patterns of library-economic disadvantage interaction (Figure \ref{fig:pirls_int}). The most obvious pattern shown in the interaction plot (Figure \ref{fig:pirls_int}) is that students from Bulgaria (blue lines) have higher literacy scores than students from Georgia (green lines). However, Bulgarian students' literacy scores also appear to decrease with increasing economic disadvantage; furthermore, the rate of decline appeared to be slower for better resourced-schools (i.e., those with more books). These patterns, expected from general socioeconomic principles, were generally observed across many countries. Georgia appears unusual:  economic disadvantage had little overall effect on literacy scores, and the effects of library size were opposite from those expected -- schools with large libraries appeared to have higher literacy scores than those with small libraries when schools were well off, but library size made little difference for strongly disadvantaged schools. 

\newpage
\section{Discussion} \label{sec:Conclusion}

In this article, we introduced a new variance-covariance structure, \texttt{rr}, in \texttt{glmmTMB}, to add reduced-rank functionality to mixed models. This feature broadens the scope of models that can be fitted by writing a large dimensional multivariate random effect as a linear combination of $d$ latent variables, a more parsimonious structure that can be more readily estimated when the dimension of the random effect is large. In Section \ref{sec:intro}, we discussed available tools in \texttt{R}, such as \texttt{gllvm}, which also fit latent variable models. These packages were developed with a primary focus on models for ecological data. The key advantage of our work is adding a factor analytic term to the suite of random effects structures already available in \texttt{glmmTMB}, such that generalized latent variable models can now be fitted to complex study designs, using a familiar interface.

We presented two applications to illustrate the use of a reduced-rank approach. In both examples the dimension of the random effect was moderate -- 9 and 15 for these two examples -- but this was already too large to be practically estimable, necessitating the use of a reduced-rank model. Reduced-rank models are capable of fitting random effects of very large dimension: for example, \citet{niku2019gllvm} fitted a GLVM with a dimension of 985 to a microbial data set. When fitting models to larger data sets, difficulties can be encountered; for example, in ecology the number of species is often large compared to the number of samples. In situations like this, it may be useful to fit a model multiple times with different starting values for the parameters, and the fit with the highest log-likelihood value is considered the best fitting model.

The reduced-rank model contributes to the model-simplification toolbox, allowing for a more parsimonious random effect which may be necessary for some study designs \citep{matuschek2017balancing}. Currently, available methods for simplification are assuming a diagonal variance-covariance matrix, with homogeneous or heterogeneous variances; assuming compound symmetry; or assuming some specific form of structure (AR(1), Toeplitz, etc.). All of these structures are available in \texttt{glmmTMB}. Another alternative for controlling complexity is some form of shrinkage towards zero on the factor loadings as proposed in a Bayesian framework \citep{bhattacharya2011sparse}.

A key step in applying a reduced-rank random effect is choosing the rank $d$. Different strategies may be used for choosing $d$, depending on the analysis goal. In the wind farm application, we used a two-dimensional biplot to visualise correlations across species (Figure~\ref{fig:wford}) and for this purpose $d=2$ was appropriate. In our second application, the PIRLS study, our goal was to make inferences about correlated fixed effects, and the reduced rank approach was used to estimate this correlation. In this case we used information criteria to choose a value for $d$ that gave us a good fit to the data. We found that estimates and confidence intervals for the fixed effect estimates were robust to different choices of rank (see supplementary material, Figure \ref{fig:wf_fixef_ds} and Figure \ref{fig:pirls_fixef_ds}). The extent to which fixed effects can change as covariance assumptions on random effects change $\boldsymbol{\Sigma}$ is a function of how much the fitted covariance structure actually changes. Factor analytical terms offer diminishing returns in terms of changes to $\boldsymbol{\Sigma}$ as $d$ increases, so there is greatest capacity for changes in interpretation when $d$ is small (in practice we have seen qualitatively important changes only for $d<2$, as in Figure~\ref{fig:pirls_fixef_ds}).

The reduced-rank structure has an interesting point of difference from other approaches to fitting a multivariate random effect in that it permits a singular variance-covariance matrix, or more precisely, it assumes singularity. Other methods of fitting correlated random effects require a positive-definite variance-covariance matrix and return warnings when encountering (near-)singularity, an issue circumvented here by assuming a reduced-rank structure. 

Our examples give just a few tastes of how reduced rank variance-covariance structures can be used in mixed modelling, where it has previously been technically difficult to fit models with random effects of high dimension. Some example areas where we see potential use for this approach include factor analysis with multi-level designs or repeated measures, and genotype-by-environment interaction analysis  \citep{piepho1997analyzing, smith2001analyzing}. We see a myriad of potential applications, and look forward to seeing how this new tool is used in practice.  

\bibliographystyle{plainnat}
\bibliography{reference}

\newpage

\appendix

\section{Model summaries for the wind farm example}\label{app:summary} 

Summary of the model for the wind farm example in Section \ref{sec:intro} fitted using \texttt{lme4} is as follows:
\begin{verbatim}
R> summary(glmer(abundance ~  Zone + Year + (Species + 0 | ID),  
+    family = "poisson", data = wf.ex))
\end{verbatim}  
\small
\begin{verbatim}
Generalized linear mixed model fit by maximum likelihood (Laplace Approximation) ['glmerMod']
 Family: poisson  ( log )
Formula: abundance ~ Zone + Year + (Species + 0 | ID)
   Data: wf.ex

     AIC      BIC   logLik deviance df.resid 
  1310.5   1336.1   -648.3   1296.5      277 

Scaled residuals: 
     Min       1Q   Median       3Q      Max 
-1.38779 -0.67079  0.06129  0.43556  1.58605 

Random effects:
 Groups Name            Variance Std.Dev. Corr 
 ID     SpeciesTorsk    0.7603   0.8719        
        SpeciesTanglake 0.9839   0.9919   -0.74
Number of obs: 284, groups:  ID, 142

Fixed effects:
            Estimate Std. Error z value Pr(>|z|)    
(Intercept)  0.63001    0.10942   5.758 8.52e-09 ***
ZoneN        0.07242    0.12807   0.565  0.57174    
ZoneS       -0.57875    0.19527  -2.964  0.00304 ** 
Year2010     0.57457    0.10360   5.546 2.92e-08 ***
---
Signif. codes:  0 ‘***’ 0.001 ‘**’ 0.01 ‘*’ 0.05 ‘.’ 0.1 ‘ ’ 1

Correlation of Fixed Effects:
         (Intr) ZoneN  ZoneS 
ZoneN    -0.343              
ZoneS    -0.525 -0.014       
Year2010 -0.493  0.057 -0.130    
\end{verbatim}

\newpage
\normalsize
The summary of the model fitted using \texttt{glmmTMB} is as follows:
\small
\begin{verbatim}
R> summary(glmmTMB(abundance ~  Zone + Year + (Species + 0 | ID), 
+    family = "poisson", data = wf.ex))
\end{verbatim}  

\begin{verbatim}
 Family: poisson  ( log )
Formula:          abundance ~ Zone + Year + (Species + 0 | ID)
Data: wf.ex

     AIC      BIC   logLik deviance df.resid 
  1310.5   1336.1   -648.3   1296.5      277 

Random effects:

Conditional model:
 Groups Name            Variance Std.Dev. Corr  
 ID     SpeciesTorsk    0.7603   0.8719         
        SpeciesTanglake 0.9839   0.9919   -0.74 
Number of obs: 284, groups:  ID, 142

Conditional model:
            Estimate Std. Error z value Pr(>|z|)    
(Intercept)  0.63000    0.10942   5.758 8.53e-09 ***
ZoneN        0.07243    0.12807   0.566  0.57170    
ZoneS       -0.57876    0.19528  -2.964  0.00304 ** 
Year2010     0.57457    0.10360   5.546 2.92e-08 ***
---
Signif. codes:  0 ‘***’ 0.001 ‘**’ 0.01 ‘*’ 0.05 ‘.’ 0.1 ‘ ’ 1
\end{verbatim}

\newpage

\normalsize
\section{Parametric bootstrap analysis for the wind farm example}\label{sec:par}

A parametric bootstrap to test the interaction terms of zone and year for the wind farm data. The P-value is estimated by comparing the observed likelihood ratio statistic to the simulated  distribution of the test statistic under the null hypothesis. Bootstrap replications which failed to converge are ignored.

\begin{Verbatim}[fontsize=\small]
R> LRobs <- 2 * logLik(wf.glmm) - 2 * logLik(wf.glmm.null)
R> library(boot)
R> lrt.fun <- function(data) {
+    library(glmmTMB)
+    null <- try(glmmTMB(abundance ~  Zone + Year + diag(Zone + Year|Species) + 
+      (1|Station) + rr(Species + 0 | ID, d = 2),
+      family = "poisson", data = data))
+    alt <- try(glmmTMB(abundance ~ Zone * Year + diag(Zone * Year|Species) + 
+      (1|Station) + rr(Species + 0 | ID, d = 2),
+      family = "poisson", data = data))
+    LR <- tryCatch({2*logLik(alt) - 2*logLik(null)}, error = function(e) {NA})
+    return(LR)
+  }
R> sim.abund <- function(data, mle) {
+    library(glmmTMB)
+    out <- data
+    out$abundance <- simulate(mle)$sim_1 #simulate data under the null
+    out
+  }
R> wf.boot <- boot(data = windfarm, 
+    ran.gen = sim.abund,
+    lrt.fun,
+    mle = wf.glmm.null,
+    sim = "parametric",
+    R = 1000,
+    parallel= "snow", #for Windows 
+    ncpus = 4)
R> p <- ( sum(wf.boot$t[,1] >= LRobs, na.rm=TRUE) +1 )/(1000 + 1)
\end{Verbatim}

\newpage
\section{Sensitivity analysis}\label{sec:sense}

\begin{figure}[H]
\includegraphics[width=\textwidth]{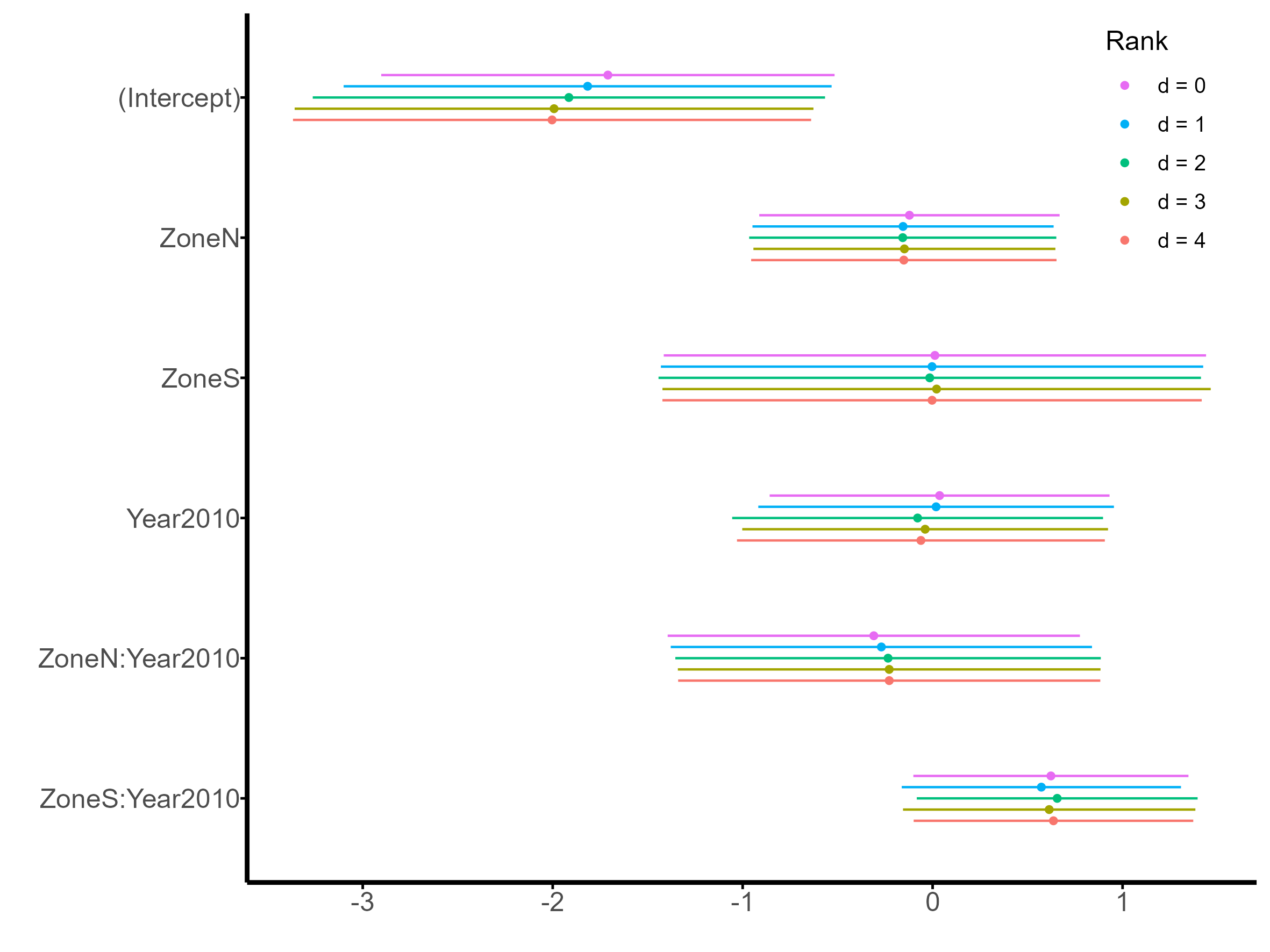}
\caption{The fixed effect estimates and 95\% confidence intervals for the wind farm model are similar when the rank (d) of the reduced-rank random effect varies from zero to four.}
\label{fig:wf_fixef_ds}
\end{figure}
\newpage
\begin{figure}[H]
\includegraphics[width=\textwidth]{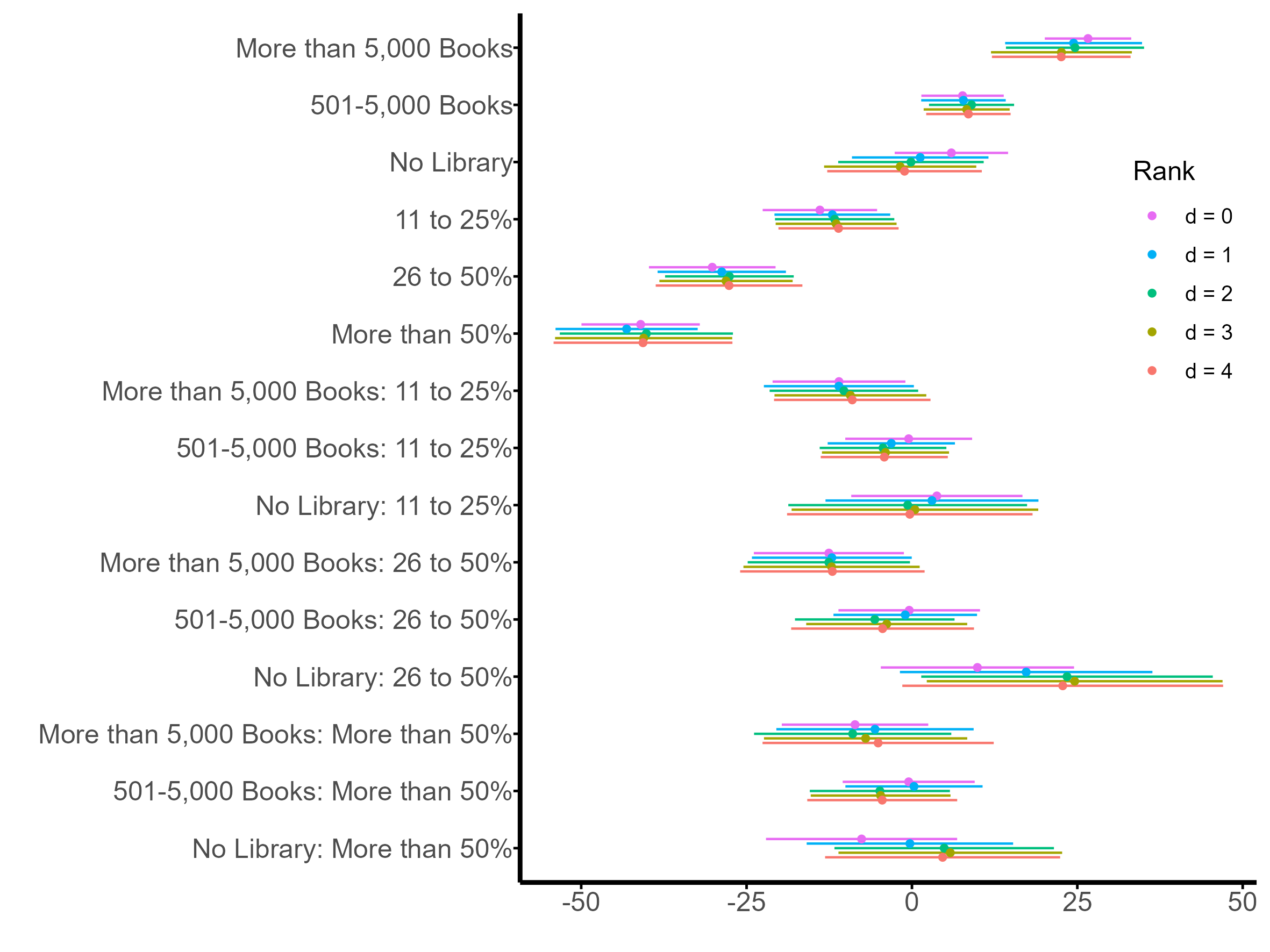}
\caption{The fixed effect estimates and 95\% confidence intervals for the PIRLS model are similar when the rank (d) of the reduced-rank random effect varies from one to four. When the reduced-rank random effect is replaced by a random intercept ($d = 0$), the estimates of the fixed effects are less similar and the standard errors may be smaller.  }
\label{fig:pirls_fixef_ds}
\end{figure}

\newpage

\section{Reduced-rank random effect estimates from the PIRLS model}
\begin{figure}[H]
\includegraphics[width=\textwidth]{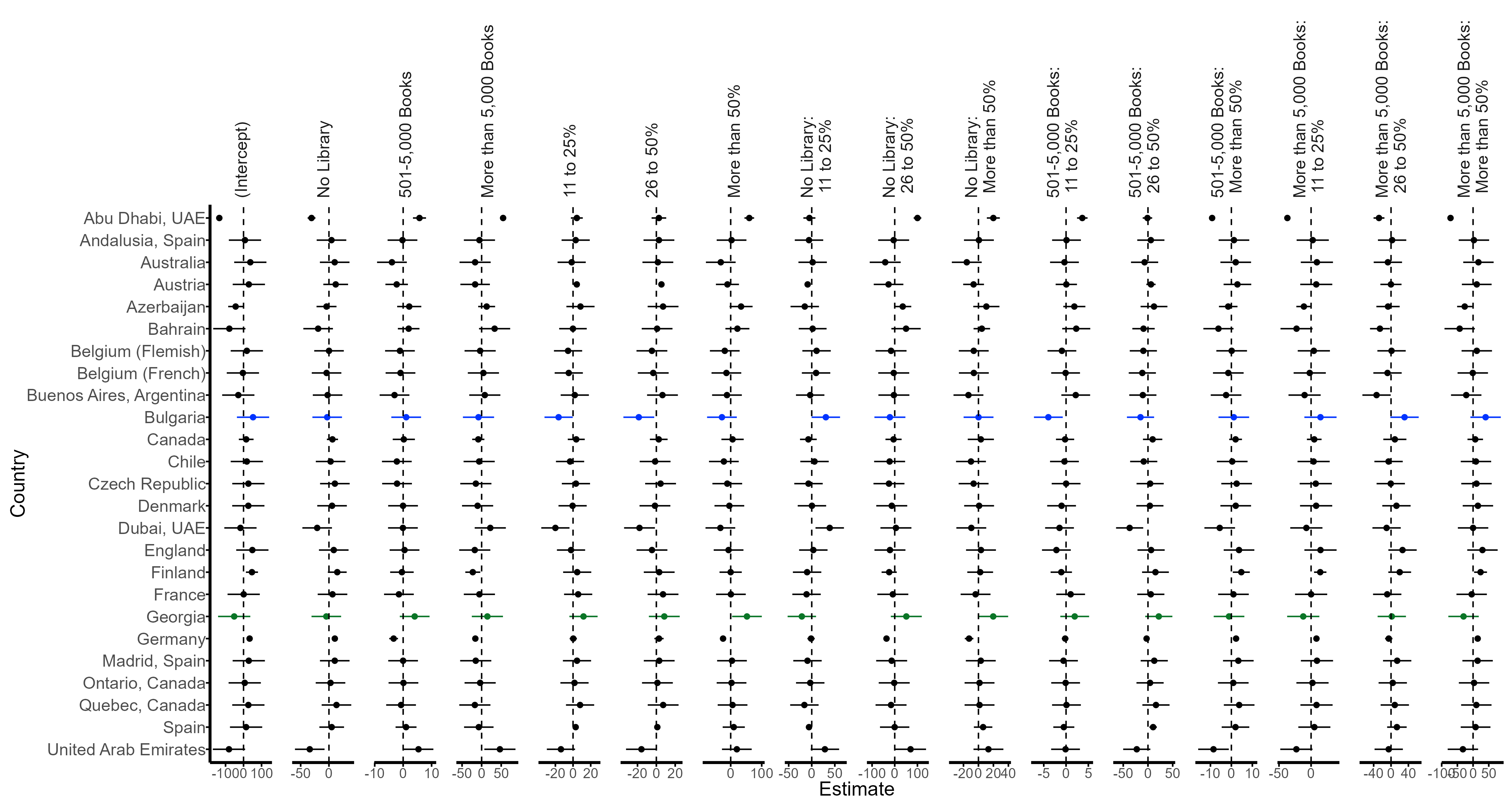}
\caption{Conditional estimates of school variables by country from the reduced-rank random effect in the PIRLS model.}
\label{fig:pirls_re}
\end{figure}
\newpage

\end{document}